# Reward driven workflows for unsupervised explainable analysis of phases and ferroic variants from atomically resolved imaging data

Kamyar Barakati,[1,a] Yu Liu,[1] Chris Nelson,[2] Maxim A. Ziatdinov,[3] Xiaohang Zhang,[4] Ichiro Takeuchi,[4] and Sergei V. Kalinin,[1, 3, b]

[1] Department of Materials Science and Engineering, University of Tennessee, Knoxville, TN 37996

[2] Center for Nanophase Materials Sciences, Oak Ridge National Laboratory, Oak Ridge, TN 37831

[3] Pacific Northwest National Laboratory, Richland, WA 99354

[4] Department of Materials Science and Engineering, University of Maryland, College Park, MD 20742

Rapid progress in aberration corrected electron microscopy necessitates development of robust methods for the identification of phases, ferroic variants, and other pertinent aspects of materials structure from imaging data. While unsupervised methods for clustering and classification are widely used for these tasks, their performance can be sensitive to hyperparameter selection in the analysis workflow. In this study, we explore the effects of descriptors and hyperparameters on the capability of unsupervised ML methods to distill local structural information, exemplified by discovery of polarization and lattice distortion in Sm-doped $BiFeO_3$ (BFO) thin films. We demonstrate that a reward-driven approach can be used to optimize these key hyperparameters across the full workflow, where rewards were designed to reflect domain wall continuity and straightness, ensuring that the analysis aligns with the material's physical behavior. This approach allows us to discover local descriptors that are best aligned with the specific physical behavior, providing insight into the fundamental physics of materials. We further extend the reward driven workflows to disentangle structural factors of variation via optimized variational autoencoder (VAE). Finally, the importance of well-defined rewards was explored as a quantifiable measure of success of the workflow.

---

[a] kbarakat@vols.utk.edu
[b] sergei2@utk.edu

**Introduction**

Physical and chemical functionalities of solids are inseparably related to the atomic structure. Phases and defects are defined based on the relative arrangement of atoms in solids. Ferroic variants arise from small symmetry-breaking distortions caused by the correlated displacement of atoms from their high-symmetry positions. The relative magnitude of these displacements is determined by the nature of primary order parameter in the system and reduces from ferroelastic to ferroelectric and ferromagnetic systems.[1] The over a hundred-year-old history of X-Ray and neutron scattering[2, 3] as the enabling element of condensed matter physics and materials science is related to its capability to probe the details of atomic structure on the macroscopically averaged level.[4, 5]

Over the last decade, the progress in aberration corrected electron microscopy[6-8] has enabled direct mapping of atomic structures with picometer level precision.[9] With this, the physical order parameter fields can be mapped on the atomic level simply from atomic positions of the cation and anion columns. The work of Jia et al. has demonstrated the direct spatially resolved mapping of polarization[10] via transmission electron microscopy (TEM),[11-13] and Chisholm et al. demonstrated this approach via Scanning Transmission Electron Microscopy (STEM).[14] Since then, this approach has been demonstrated for exploration of other order parameters including octahedra tilts[15-18] and chemical expansivity,[19] and a number of studies reporting the atomically resolved structures of charged and uncharged topological defects[18, 20, 21], order parameter coupling at interfaces[22, 23], modulated phases,[24] polarization pinning at the defects[5], exotic structures such as ferroelectric vortices[25] and many other aspects of materials behavior.[21, 26] In several cases, these studies have been used to reconstruct the associated physical mechanisms including vacancy-induced polarization screening[19, 27] or parameters of associated Ginzburg-Landau models.[28, 29]

In addition to physics-based analysis of the STEM data, several approaches based on the machine learning methods have been reported.[30, 31] In these cases, the structure is represented using descriptors that capture the characteristics of the local structure. The descriptors can be coordinates of the near neighbourhood,[32] image patch centered on the specific atoms,[33] or image patches centered on the grid.[34] The clustering or dimensionality reduction of the descriptors yields the class labels or reduced descriptors that can be mapped on the original image space, illustrating the variability of local structures.[35, 36] These in turn can be used to visually identify ferroelectric domains[37-39], single phase regions[40, 41], phase intergrowth[42, 43], topological and structural defects[44, 45], and other relevant structural motifs.

Both physics- and ML-based analysis of the imaging data can be represented as a workflow, or sequence of image analysis operations including background corrections[46, 47], key point (e.g. atom) finding, descriptor construction, clustering, and dimensionality reduction.[48] This process is now driven by a human operator and is extremely time-consuming, often taking days and weeks. Correspondingly, only a small fraction of microscope data is analyzed. Furthermore, the analysis process is often non-myopic in that the selection of descriptors early in the analysis pipeline can affect the data representations down the line, resulting in strong human bias in analysis. Finally, the extreme time requirements and human orchestration make this approach inconsistent with real-time data analytics and especially autonomous microscope operations.[49]

Recently, we have proposed an approach for construction of data analysis pipelines based on the reward-driven workflow concept.[50] In these, human operator selects a set of reward functions that in some way reflect the quality of the analysis results. These reward functions are designed to balance the human heuristics and physics of the system. Based on this reward function, the construction of the data analysis pipeline can be represented as an optimization problem in the combinatorial space of possible operation sequences and product space of hyperparameters for each operation. In this manner, the workflow construction is reduced to the optimization problem in the high-dimensional space, and as such is robust, explainable, and unsupervised. While ostensibly biased, this approach in fact directly matches the decision making of the human operator, the biases introduced by the introduction of reward function are traceable, and reward functions can be tuned during the experiments.

Previously, we have demonstrated the reward driven analysis approach for the atom finding in the post-acquisition[50] and real-time setting.[51] Here, we demonstrate how this concept can be applied for the segmentation of the atomically–resolved STEM images towards identification of phase and ferroic variants.

I.  **Model systems**

The dataset for our model system includes cross-sectional atomic resolution STEM images of Sm-doped $BiFeO_3$ (BFO) at nominal Sm concentrations of 0%, 7%, 10%, 13%, and 20%. Samples were prepared using focused ion beam (FIB)[52] extraction from a single composition-spread film. Pure $BiFeO_3$ exists as a rhombohedral ferroelectric (R3c) phase, however, when small amount of Bi is replaced with Sm, the material transits into an orthorhombic non-ferroelectric phase. These two distinct phases are separated by a morphotropic phase boundary (MPB)[53], characterized by incompatible symmetries. Currently, this dataset contains the original images, atomic coordinates, intensities for atomic columns, and calculated physical order parameter fields.[54] As such, these can be used both for development of the new analysis pipelines, benchmarking the analysis results, anomaly detection and physics discovery for the morphotropic materials.

Here, we develop the reward driven workflow for the phase and ferroic variant identification based on the atomically centered descriptors. In this, the coordinates for the atomic columns are assumed to be known, and the analysis starts with the filtering, selection of the descriptors as image patches, and subsequent clustering and dimensionality reduction. The output of such analysis are the real space representations of labels and corresponding centroids (for clustering) and latent variables (for dimensionality reduction) that would ideally separate phases and ferroic variants. Previously, we have demonstrated this approach for ad-hoc workflows with human (and very time consuming) tuning of the descriptors and hyperparameters.[35] Here, we cast it as an unsupervised analysis problem.

## II. Phase and ferroic variant identification

First, we explore the descriptor and hyperparameter optimization for clustering-based workflows. Here, we identify the descriptor as rectangular window of the size ($w_1$, $w_2$) centered on the atomic columns with coordinates ($x_i$, $y_i$), where $i = 1, \ldots, N$ is the total number of atomic columns in the image. Atoms in the chosen sublattice act as key points, and sub-images centered on these atoms are used as descriptors. By using experimentally determined atomic coordinates instead of assuming an ideal periodic lattice, this method accounts for local strains and distortions, providing a more accurate representation of the material's microstructure and its influence on the order parameter. The formed image patches are represented as $w_1 w_2$ dimensional vectors and are classified using Gaussian Mixture Model (GMM).[55, 56] The corresponding class labels are plotted as the maps centered on ($x_i$, $y_i$) as shown in Figure 1. The centroids for the GMM clusters are available in the associated Colab notebook.

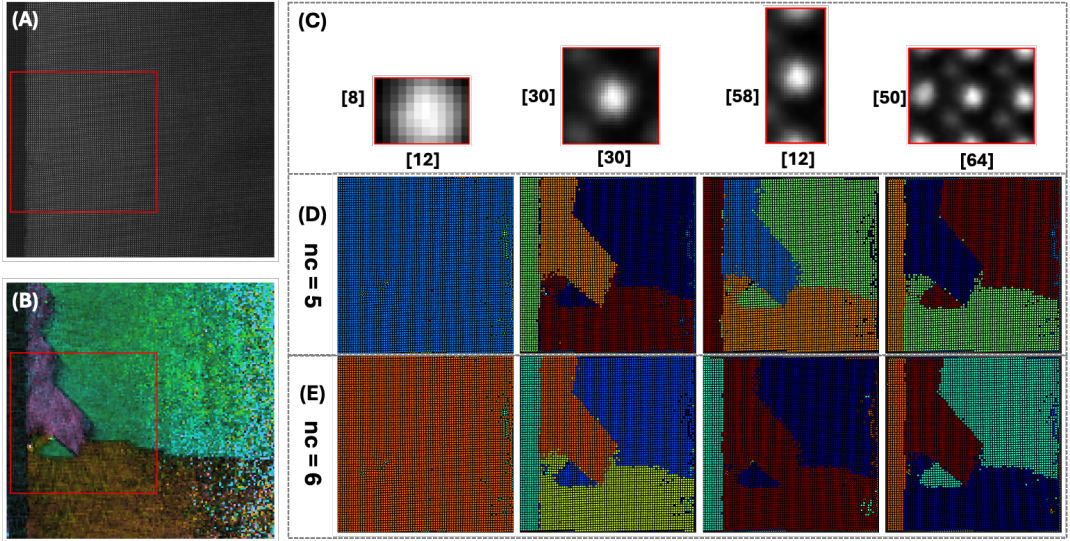

**Figure 1**: (A) HAADF image showing the selected region of interest used for analysis. (B) Ground truth polarization map of the region, illustrating the actual polarization distribution. (C) Selected descriptor examples. (D) Clustered regions using GMM with 5 fixed components, and (E) clustered regions using GMM with 6 fixed components, both illustrating the effects of window size selection and cluster count on segmentation accuracy and granularity.

Figure 1 illustrates how different window sizes influence region segmentation in a selected area of Sm-doped BiFeO$_3$ thin films using GMM clustering. These descriptors encode local structural information, including polarization and lattice distortions, and GMM clusters regions with similar structural properties by representing these variations as a combination of Gaussian distributions.[57] The identification of distinct clusters corresponds to different phases or domains in the material. Smaller descriptors, such as (8, 12), tend to over-segment the image, detecting minor variations and introducing noise. This can obscure the physical meaning of the domains by over-emphasizing local fluctuations in the material. In contrast, larger window sizes, such as (50, 64), provide more generalized segmentation by averaging over larger areas, which can result in the loss of fine details crucial for accurately identifying domain boundaries. On the other hand, the number of GMM components directly affects how the image is segmented into different phases. With fewer components, the clustering captures broader, larger-scale regions, while increasing the number of components allows for finer distinctions between different phases. However, an excessively large number of components may lead to overfitting, where minor variations in the descriptors are misinterpreted as distinct phases. Hence, the descriptor and number of GMM components must be carefully selected to achieve a balance between capturing the physical meaning of the underlying domain structures and avoiding over-segmentation or loss of important details.[58]

To quantify the effect of window size selection on clustering results, we applied k-means clustering[59, 60] to convert the continuous ground truth $P_{xy}$ polarization map into discrete labels. This allowed for a direct comparison of labels between the ground truth and the images generated by GMM clustering across different window sizes. For each combination of window sizes, we calculated the correlation coefficient[61] between the ground truth labels and the GMM-derived labels. The heat maps for each parameter were then combined into a single RGB heat map to visualize how window size influences clustering alignment with the ground truth polarization states.

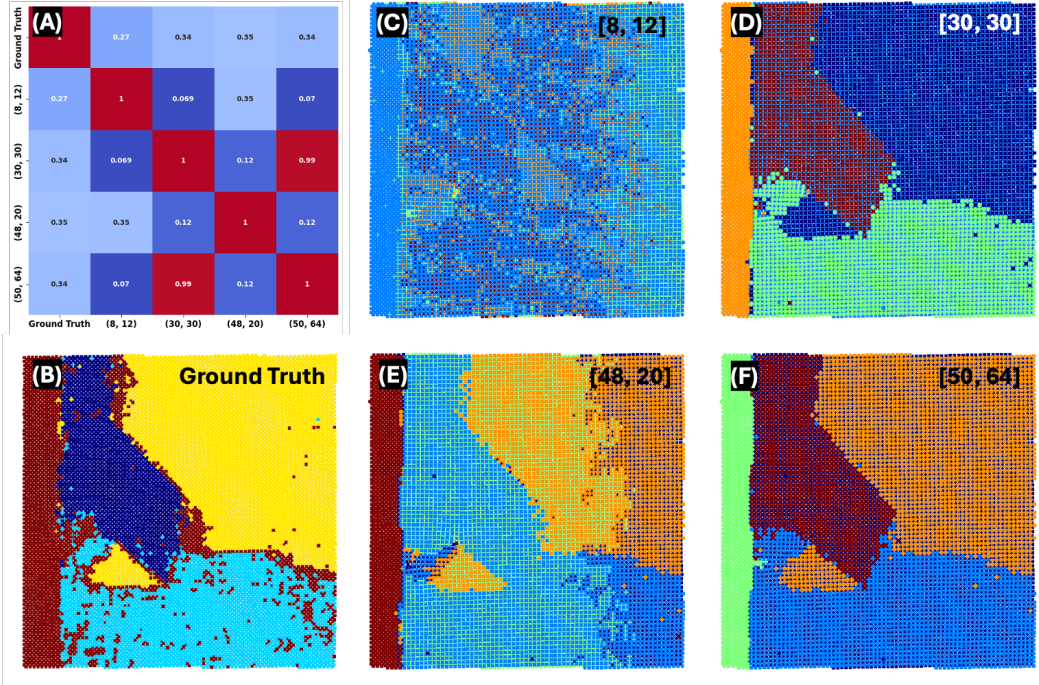

**Figure 2:** A) Heat map of the correlation coefficient between ground truth images of polarization component $P_{xy}$ with the label images for the pairs of ($w_1$, $w_2$), B) Ground truth labeling with color-coded polarization regions, C) to F) Segmentation of regions based on the descriptors sizes and fixed 5 number of components in GMM clustering.

Figure 2 illustrates the inherent complexity of selecting optimal hyperparameters, particularly window sizes, within the workflow. Figure 2(A) displays the correlation heatmap between various window sizes and the ground truth. Notably, none of the selected window sizes by human operator perfectly align with the ground truth values, indicating that while we may infer one parameter (e.g., window size), other hyperparameters may critically influence the results.

Figure 2(C-F) further emphasizes this point by depicting the spatial distributions of different window sizes [(8, 12), (30, 30), (48, 20), and (50, 64)]. These variations reveal that each window size captures distinct structural features, but no single window size fully reproduces the true material characteristics shown in the ground truth image Figure 2(B). The discrepancies between the various window sizes demonstrate that incorrect or suboptimal selection of window

size may lead to significant variations in the detected structural features, such as domain boundaries or polarization patterns. It is remarkable that changing the descriptor size visualizes different aspects of domain pattern, highlights the domain walls, or indicates presence of the extraneous phases. Also note that in Figure 2(E) for this particular descriptor a new cluster is visualized, corresponding to the effect of the mistilt on the right hand side of the image.[62]

This analysis illustrates a very complex behavior of unsupervised analysis even in the 2D parameter space of simple clustering algorithm. Practically, the dimensionality of the parameter space for the full workflow is considerably higher, and includes parameters of denoising process, window sizes, selection of the clustering algorithm per se, and the hyperparameters of clustering. For instance, in a workflow involving a Gaussian blur filter[63] and GMM clustering, the parameter space can include both continuous and categorical variables. The continuous parameters may consist of the standard deviation (sigma, $\sigma$) of the Gaussian blur filter and the dimensions of the descriptor window (width and height). On the other hand, the categorical variables may include the number of clusters, the covariance type in GMM (full, tied, diagonal, or spherical), and the selection of descriptors (e.g., image patches or atomic coordinates). Selection of alternative clustering methods, image transform or linear or non-linear dimensionality reduction prior to clustering further increases the parameter space dimensionality.

To optimize the analysis workflow and make it traceable and robust, we employ the reward-driven concept. Here, we utilize the fact that in ferroelectric materials the domain walls are continuous, tend to align in certain crystallographic dimension, and have relatively small roughness. To identify the wall geometry, we apply an edge-detection algorithm[64] to identify the boundaries between the GMM clusters effectively mapping the domain walls. To represent these boundaries as lines, we employed a line detection method, specifically the Hough Transform[65, 66], ensuring that the detected domain walls were accurately fitted as continuous line segments. With access to this line-based representation of domain walls, we proceeded to optimize the analysis through a reward-driven approach.

Two rewards were defined to guide the optimization process within a parameter space consisting of three key hyperparameters: window size $w_1$, window size $w_2$, and the GMM covariance type. Reward_1 focuses on minimizing the curvature of the detected domain walls, encouraging the formation of smooth and continuous lines that accurately represent the natural morphology of the walls without abrupt discontinuities. Physically, ferroelectric domain walls are expected to exhibit low curvature to maintain stability and minimize elastic energy.[67] The curvature at each point was computed using the gradients of the x and y coordinates, followed by their second derivatives. Specifically, the curvature κ was computed using the formula:

$$\kappa = (|d_x \cdot d_y^2 - d_y \cdot d_x^2|/(d_x^2 + d_y^2))^{1.5} \qquad (1)$$

where $d_x$ and $d_y$ the gradients of the x and y coordinates, and $d^2x$ and $d^2y$ are their respective second derivatives. This ensures that any abrupt changes in direction, indicating high curvature, are penalized. We then calculated the average curvature across all points and minimized this value. This drives optimization toward smoother lines, reflecting the natural configuration of ferroelectric domain walls. Reward_2 aims to maximize the length and continuity of the detected domain walls. This reward is rooted in the physical expectation that ferroelectric domain walls are extended structures that typically form continuous patterns without segmentation. The goal is to avoid detecting fragmented or overly segmented domain walls, thereby improving the representation of the physical reality. We calculated the total length of each segment using the Euclidean distance[68] between consecutive points within the segment. The reward for each cluster of segments was then computed as the total length of all segments divided by the number of segments in the cluster, effectively penalizing excessive segmentation while encouraging longer, continuous walls.

$$L = (total\ length^{(i)} / \#\ of\ segments^{(i)}) \qquad (2)$$

Together, these two rewards guided the optimization process to produce results that are both physically meaningful and consistent with the natural morphology of ferroelectric domain walls.

Defining appropriate rewards in an optimization workflow, particularly in experimental sciences, is inherently challenging. The difficulty lies in quantifying experimental goals while ensuring that these rewards convey the physical meaning of our objectives. For ferroelectric domain walls, this involves capturing features like curvature, continuity, and morphology, which reflect fundamental physical properties. The process of defining these rewards may vary depending on the individual's perspective and understanding of the experimental context. By establishing rewards that embody the same core principles while allowing for different interpretations, we ensure that the workflow remains adaptable yet converges to scientifically consistent outcomes.

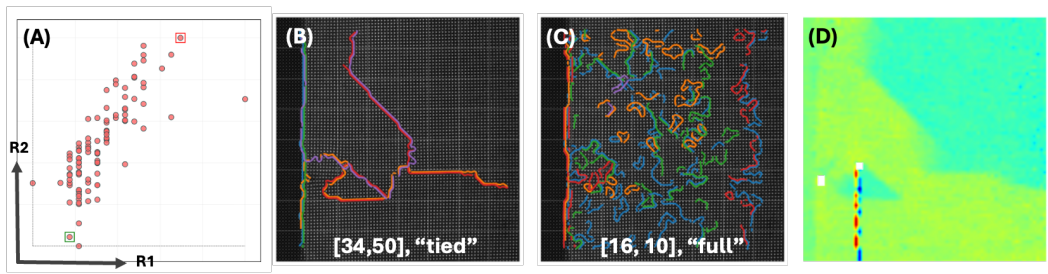

**Figure 3:** (A) Pareto front solutions representing the trade-off between two rewards, $R_1$ (Straightness) and $R_2$ (Length). The red and green markers highlight the solutions selected by operator during analysis according to be prerequisites of the experiment, (B) Solution acquired by the workflow, showing the best possible trade-off between objectives of the experiment, (C) Another option in pareto front solution that operator has selected to explore the materials properties, (D) Ground truth polarization map.

The reward-driven workflow produced 100 different solutions, each exploring various combinations of hyperparameters to satisfy the analysis objectives. In Figure 3(A), each point corresponds to three parameters: the height and width of the window size, and the GMM clustering hyperparameters (in this case, the covariance type). The green point in Figure 3(A) represents the optimal solution with the highest rewards. This is shown in Figure 3(B), where both predefined objectives straightness, continuous lines with minimal curvature have been successfully achieved, considering a descriptor of size [34, 50] and covariance type of "tied" as hyperparameters of the workflow. The red point in Figure 3(A) represents a selected solution chosen for exploration. The corresponding predicted pattern, shown in Figure 3(C), demonstrates that both rewards are at their minimum. This outcome indicates that neither of the defined objectives was successfully achieved.

Note that moving across the possible solution space reveals different aspects of the domain structure and allows to visualize the domain walls and the boundary of the mis-tilt region. Importantly, this decision making can be performed as a part of real-time instrument operation.

### III. Disentangling the factors of variation

As a second example of reward driven unsupervised workflow, we explored the disentanglement of the structural factors of variation using the variational autoencoder (VAE) based approach.[69, 70] The VAE offers a non-linear dimensionality reduction of the descriptors, disentangling the factors of variation within the data. In many cases, these factors of variation can be identified with the physical and chemical variability in the system. For this analysis, we used the pyroVED package developed by Ziatdinov *et al*.[71] Recent review of VAE approach is reported in Valleti *et al*.[72]

To explore the physics of materials, we use conditional rotationally invariant Variational Autoencoder (CrVAE), incorporating several degrees of freedom. Two latent variables were used to capture intrinsic features of the material, while rotational transformations accounted for symmetry in the data. The conditional variable is the atom type, i.e. patches centered at the A site and B site of the atoms in material's lattice structure. The A site is mainly occupied by $Sm^{3+}$ (Samarium ions) and $Bi^{3+}$ (Bismuth ions), which sit at the corners of the unit cell. The B site is occupied by $Fe^{3+}$ (Iron ions), located at the center of the unit cell and coordinated by oxygen atoms to form $FeO_6$ octahedra. These conditional dimensions enable the model to focus on specific structural features. In Figure 4(A), the smooth transitions observed in the latent dimensions across the A site cations suggest subtle shifts in lattice stability and polarization, likely driven by minor variations in the local chemical environment and structural properties of the material. Conversely, in Figure 4(B), the transitions at the B site reveal more pronounced structural changes. These are associated with the greater flexibility of the $Fe^{3+}$ ions and the $FeO_6$ octahedra, reflecting alterations in ferroelectric and magnetic properties, as well as lattice strain.

The correlation between the representation of unit cell centers along the $z_1$ - $z_2$ axis in Figure 4(C) and the phase transition from ferroelectric to non-ferroelectric states shown in Figures 4(D, E) indicates the existence of more than two distinct structural variations within the material. Note that for ideal material we expect that polarization variants will be discovered to correspond to the same value of latent variable but different orientations. However, for realistic images the mistilt effect break this rotation invariance, and ferroic variants are separated as different latent values even for rVAE. The rotational latent angle now visualizes the polarization rotation at the domain walls, as clearly visible in Figure 4(F).

To assess the reliability of the model's predictions, we computed the latent uncertainty across different regions of the material by analyzing the variance of multiple dropout samples. This technique provides a measure of how confidently the model can reconstruct material features based on its training data. High uncertainty regions indicate areas where the model's decoder is less reliable, due to sparse data representation or complex structural transitions, such as defects or phase boundaries. In contrast, regions with low uncertainty suggest a stable and confident reconstruction of the material's atomic structure. The latent uncertainties of $\sigma_1$ - $\sigma_2$ and angular orientation in Figures 4(J), 4(K), and 4(L) highlight how these encoded features vary spatially across the material, with the uncertainty map serving as a key tool to identify areas where further refinement or additional data may be needed.

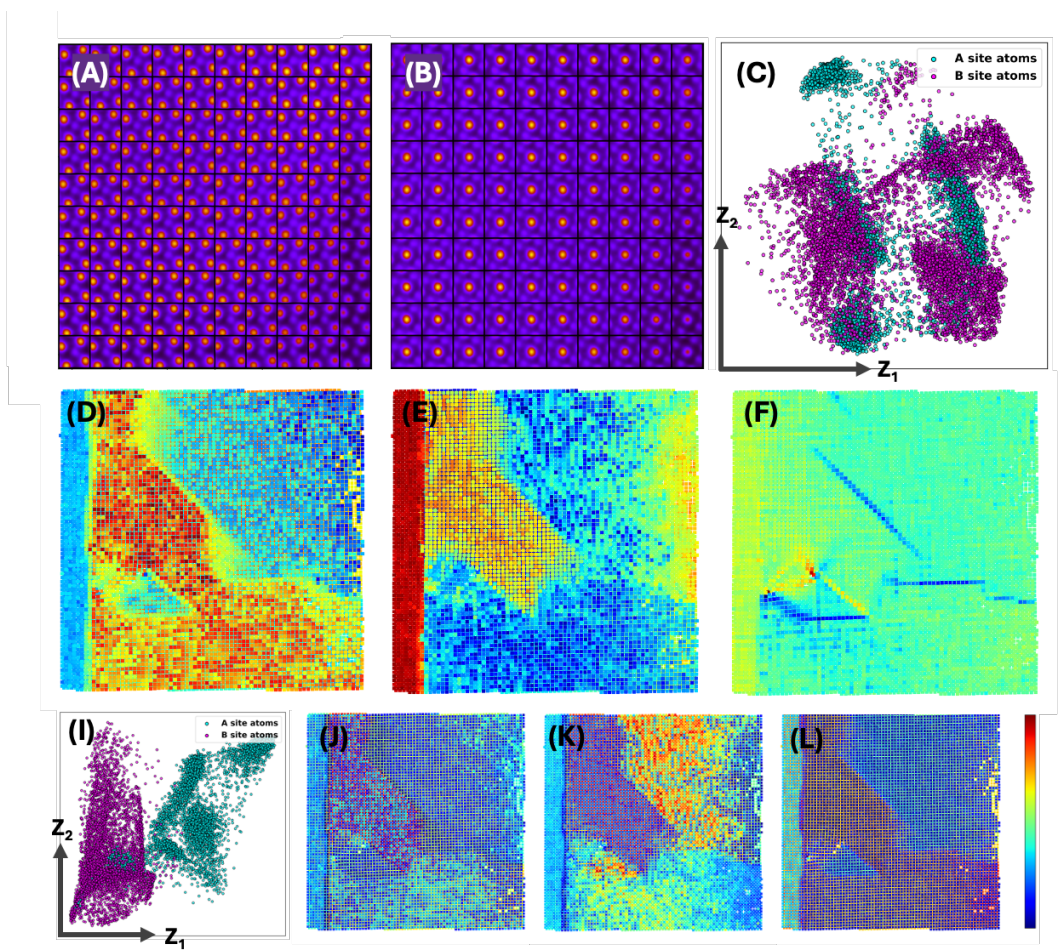

**Figure 4**: Conditional rotational variational autoencoder considering possible degree of freedom for a certain patch size of (40, 40), (A) 2D grid visualization showing the decoded images generated by the CrVAE across the latent space, where (A) and (B) show latent space transitions for A and B site atoms, respectively. (C) maps unit cell centers along the $z_1$ - $z_2$ axis, indicating distinct structural variations. (D) and (E) highlight phase transitions between different ferroelectric domains, while (F) demonstrates angular variations with $\pi/2$ rotations, consistent with ferroic domain walls. (I) represents the latent uncertainty in the material structure, derived from the two key latent variables $\sigma_1$ and $\sigma_2$. (J), (K), and (L) represent the spatial distribution of the latent uncertainty for $\sigma_1$ - $\sigma_2$, and angular orientation variables, respectively.

The VAE application towards imaging data also creates workflows with multiple tunable parameters. These include denoising and descriptor selection similar to clustering, and a broad range of hyperparameters for VAE including the invariance (rotation, translation, scale), priors on the corresponding distributions, and classical hyperparameters on the encoder and decoder including number of layers, activation functions, etc. Here, we focus discussion only on the physics relevant hyperparameters including invariances, that jointly with description form the optimization space for the reward driven workflow.

We also note that the selection of hyperparameters in VAE can also be based on reward functions. For example, classification of phases and defects calls for normal VAE, while mapping the rotation of the polarization field selects rVAE. In the case of CrVAE, the model is applied when both rotational transformations and specific known conditions, such as lattice site information (A and B sites), need to be considered. However, this determination is done prior to workflow selection, and corresponds to the degenerate case of reward function. Here, the embedded CrVAE in the reward-driven workflow follows a similar logic as before, where the optimization is guided by two key rewards. The first emphasizes consistency in the orientation, and straightness of the domain walls, reflecting the material's crystallographic symmetry and minimizing variations in angular alignment. The second focuses on ensuring continuous length representation of domain walls, avoiding any abrupt discontinuities that could distort the natural morphology.

The reward-driven workflow was employed to optimize the descriptor size for each degree of freedom in the CrVAE processing system. In this analysis, the second latent dimension ($z_2$) was used to identify in-plane domain walls by applying Kernel Density Estimation (KDE)[73] and peak detection.[74] KDE was employed to estimate the distribution of $z_2$ revealing regions of high data density that correspond to significant structural features. Peaks in the KDE were identified as potential markers for domain walls. To focus on the most relevant features, we selected the top 5 peaks with the highest densities and classified each $z_2$ value based on its proximity to these peaks. This approach allows for an effective classification of the latent space, capturing transitions in the material's structure and enabling a more detailed analysis of in-plane domain wall behavior. As shown in Figure 5(A), where Pareto-optimal points illustrate the trade-off between straightness and lines length, the goal is to achieve a solution closest to the origin [0, 0], representing optimal rewards in both objectives. Figure 5(B) shows that both objectives straightness and continuous length were successfully achieved with a descriptor size of [50, 58]. Compared to the ground truth in Figure 5(D), this analysis reveals additional insights beyond just polarization, providing a deeper understanding of the material properties. A corresponding predicted pattern for exploration is shown in Figure 5(C). Although this solution does not fully optimize the rewards, it captures some similar features. The lines representing the domain walls in this case are either inaccurately placed or may be highlighting other characteristics, such as variations in material composition or the effects of imaging conditions.

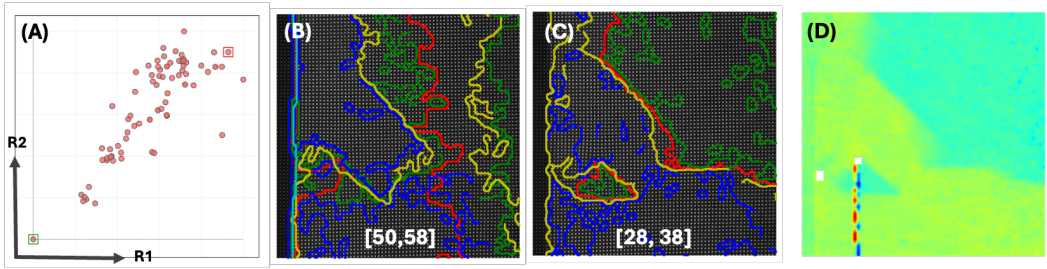

**Figure 5**: (A) Pareto front solutions representing the trade-off between two rewards, $R_1$ (straightness) and $R_2$ (length) pure $BiFeO_3$. The red and green markers highlight the solutions selected by operator during analysis according to be prerequisites of the experiment, (B) Solution acquired by the workflow, showing the best possible trade-off between objectives of the experiment, (C) Another option in pareto front solution that operator has selected to explore the materials properties, (D) Ground truth polarization map.

## IV. Evolution of ferroelectricity across composition series

To enhance our understanding of the domain structure, we applied the CrVAE embedded reward-based model on 7% Sm concentration in Sm-doped $BiFeO_3$ (BFO) to investigate its influence on domain wall formation and ferroelectric behavior. This model can be readily applied to other Sm concentrations, including 10%, 13%, and 20%, as demonstrated in the supporting notebook. This approach allows us to dynamically optimize descriptor sizes and latent representations for detecting domain walls while the material's composition changes.

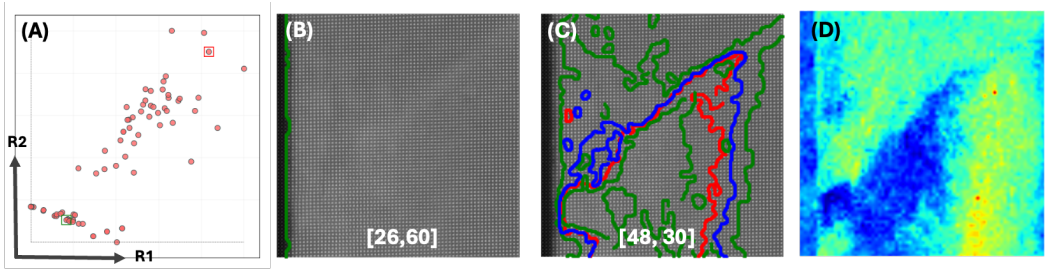

**Figure 6**: (A) Pareto front solutions representing the trade-off between two rewards, $R_1$ (straightness) and $R_2$ (length) for 7% concentration in Sm-doped $BiFeO_3$. The red and green markers highlight the solutions selected by operator during analysis according to be prerequisites of the experiment, (B) Solution acquired by the workflow, showing the best possible trade-off between objectives of the experiment, (C) Another option in pareto front solution that operator has selected to explore the materials properties, (D) Ground truth polarization map.

The data presented in Figure 6 demonstrates that the workflow effectively optimizes the selection of descriptor parameters to meet the objectives of minimizing curvature and maximizing the wall length. Although the model did not identify domain walls in this material, this outcome is aligned with our predefined criteria, which prioritized domain walls that are both straight and extended. The observed domain structures in this material, however, do not exhibit these characteristics at the current concentration level. Consequently, while the workflow effectively fulfills the specified reward conditions, this result highlights a limitation in the reward definitions when applied to more complex materials, as exemplified by the structure observed here. This suggests a need for refining the rewards criteria to better capture the natural variability and complexity of domain wall structures in MPB compositions.

To address this need, we propose creating a table of reward functions that align closely with the physical parameters and objectives of the experiment, providing a structured approach to benchmarking them. This process begins by defining reward functions that represent realistic physical expectations for domain wall structures. We then utilize crowdsourcing to gather diverse insights and establish a comprehensive list of reward functions that capture essential attributes of these structures. This approach is based on the principle that, ideally, domain walls should be short to minimize the material's free energy and align linearly with crystallographic directions. However, practical challenges complicate this ideal: we lack direct access to the free energy functional, and real materials frequently exist in non-equilibrium states with complex long-range interactions and fields. In response to these complexities, our method includes iterative benchmarking, refinement through collective expert feedback, and a curated list of functions that quantify key characteristics relevant to the study. By leveraging these targeted reward functions, we aim to achieve a nuanced representation of domain wall structures that respects both theoretical ideals and the inherent variability in real-world materials.

**Summary**

To summarize, we propose and implement reward-driven workflows for the unsupervised rapid image analysis of atomically resolved data towards the discovery of phases and ferroic variants. With atomic column coordinates known, this problem is equivalent to clustering or dimensionality reduction in the high-dimensional space of descriptors, with the analysis results being sensitive to the choice of descriptors and the hyperparameters of the unsupervised method. We cast this problem as an optimization process in the parameter space of descriptor size and hyperparameters, with the reward function representing the physics-based heuristics of the problem, aligning with human operator decision-making. For clustering, this approach allows robust and explainable segmentation of the image, which shows excellent agreement with physics-based ground truth analysis. For VAE this methodology facilitates efficient dimensionality reduction and the extraction of meaningful features that align with physical interpretations. For CrVAE, added conditions focus on crystallographic sites of atoms, extending unsupervised learning to reveal how atomic arrangements influence phase stability, and structural variations.

A unique aspect of this reward-based segmentation is that it is unsupervised, robust, and explainable. As such, it can be deployed as part of real-time data analytics on automated microscopes to support human decision-making and yield physically explainable results. We note that the reward workflow concept can be further generalized to a much broader range of physics discovery tasks if physics- or human heuristic-based reward functions can be formulated.

**Acknowledgements**

This work (workflow development, reward-driven concept) was supported (K.B., Y.L., and S.V.K.) by the U.S. Department of Energy, Office of Science, Office of Basic Energy Sciences Energy Frontier Research Centers program under Award Number DE-SC0021118. STEM imaging was performed at the Oak Ridge National Laboratory's Center for Nanophase Materials Sciences (CNMS). The work at the University of Maryland was supported in part by the National Institute of Standards and Technology Cooperative Agreement 70NANB17H301 and the Center for Spintronic Materials in Advanced infoRmation Technologies (SMART) one of centers in nCORE, a Semiconductor Research Corporation (SRC) program sponsored by NSF and NIST.

**AUTHOR DECLARATIONS**

Conflict of Interest: The authors have no conflicts to disclose.

**Author Contributions:**

Kamyar Barakati: Conceptualization (equal), Data curation (lead), Formal analysis (equal), Writing – original draft (equal), Software (equal), Methodology (equal); Sergei V. Kalinin: Conceptualization (equal), Formal analysis (equal), Funding acquisition (equal), Writing – review & editing (equal), Supervision (equal); Yu Liu: Formal analysis (equal), Investigation (equal); Maxim A. Ziatdinov: Software (equal), Investigation (equal); Chris Nelson: Data curation (equal), Xiaohang Zhang, Ichiro Takeuchi: Investigation (equal).

**DATA AVAILABILITY:**

The code supporting the findings of this study is publicly accessible on GitHub at [GitHub]

# References

(1) Nagarajan, V.; Roytburd, A.; Stanishevsky, A.; Prasertchoung, S.; Zhao, T.; Chen, L.-Q.; Melngailis, J.; Auciello, O.; Ramesh, R. Dynamics of ferroelastic domains in ferroelectric thin films. *Nature materials* **2003**, *2* (1), 43-47.
(2) Feigin, L.; Svergun, D. I. *Structure analysis by small-angle X-ray and neutron scattering*; Springer, 1987.
(3) Sinha, S.; Sirota, E.; Garoff; Stanley, H. X-ray and neutron scattering from rough surfaces. *Physical Review B* **1988**, *38* (4), 2297.
(4) Lines, M. E.; Glass, A. M. *Principles and applications of ferroelectrics and related materials*; Oxford university press, 2001.
(5) Xu, Y. *Ferroelectric materials and their applications*; Elsevier, 2013.
(6) Krivanek, O.; Dellby, N.; Spence, A.; Camps, R.; Brown, L. Aberration correction in the STEM. In *Electron Microscopy and Analysis 1997, Proceedings of the Institute of Physics Electron Microscopy and Analysis Group Conference, University of Cambridge, 2-5 September 1997*, 2022; CRC Press: pp 35-40.
(7) de Jonge, N.; Houben, L.; Dunin-Borkowski, R. E.; Ross, F. M. Resolution and aberration correction in liquid cell transmission electron microscopy. *Nature Reviews Materials* **2019**, *4* (1), 61-78.
(8) Erni, R. *Aberration-corrected imaging in transmission electron microscopy: An introduction*; World Scientific Publishing Company, 2015.
(9) Yankovich, A. B.; Berkels, B.; Dahmen, W.; Binev, P.; Sanchez, S. I.; Bradley, S. A.; Li, A.; Szlufarska, I.; Voyles, P. M. Picometre-precision analysis of scanning transmission electron microscopy images of platinum nanocatalysts. *Nat. Commun.* **2014**, *5*, 4155, Article. DOI: 10.1038/ncomms5155 From NLM PubMed-not-MEDLINE.
(10) Rodriguez, B. J.; Jesse, S.; Alexe, M.; Kalinin, S. V. Spatially resolved mapping of polarization switching behavior in nanoscale ferroelectrics. *Advanced Materials* **2008**, *20* (1), 109-114.
(11) Jia, C. L.; Urban, K. W.; Alexe, M.; Hesse, D.; Vrejoiu, I. Direct observation of continuous electric dipole rotation in flux-closure domains in ferroelectric Pb(Zr,Ti)O(3). *Science* **2011**, *331* (6023), 1420-1423, Article. DOI: 10.1126/science.1200605 From NLM PubMed-not-MEDLINE.
(12) Jia, C. L.; Mi, S. B.; Urban, K.; Vrejoiu, I.; Alexe, M.; Hesse, D. Effect of a single dislocation in a heterostructure layer on the local polarization of a ferroelectric layer. *Phys Rev Lett* **2009**, *102* (11), 117601, Article. DOI: 10.1103/PhysRevLett.102.117601 From NLM PubMed-not-MEDLINE.
(13) Jia, C. L.; Mi, S. B.; Urban, K.; Vrejoiu, I.; Alexe, M.; Hesse, D. Atomic-scale study of electric dipoles near charged and uncharged domain walls in ferroelectric films. *Nat. Mater.* **2008**, *7* (1), 57-61, Article. DOI: 10.1038/nmat2080 From NLM PubMed-not-MEDLINE.
(14) Chisholm, M. F.; Luo, W.; Oxley, M. P.; Pantelides, S. T.; Lee, H. N. Atomic-scale compensation phenomena at polar interfaces. *Phys Rev Lett* **2010**, *105* (19), 197602, Article. DOI: 10.1103/PhysRevLett.105.197602 From NLM PubMed-not-MEDLINE.
(15) Borisevich, A. Y.; Chang, H. J.; Huijben, M.; Oxley, M. P.; Okamoto, S.; Niranjan, M. K.; Burton, J. D.; Tsymbal, E. Y.; Chu, Y. H.; Yu, P.; et al. Suppression of octahedral tilts and associated changes in electronic properties at epitaxial oxide heterostructure interfaces. *Phys. Rev. Lett.* **2010**, *105* (8), 087204. DOI: 10.1103/PhysRevLett.105.087204 From NLM PubMed-not-MEDLINE.
(16) Kim, Y. M.; Kumar, A.; Hatt, A.; Morozovska, A. N.; Tselev, A.; Biegalski, M. D.; Ivanov, I.; Eliseev, E. A.; Pennycook, S. J.; Rondinelli, J. M.; et al. Interplay of octahedral tilts and polar order in BiFeO3 films. *Adv Mater* **2013**, *25* (17), 2497-2504, Article. DOI: 10.1002/adma.201204584 From NLM Medline.
(17) Borisevich, A.; Ovchinnikov, O. S.; Chang, H. J.; Oxley, M. P.; Yu, P.; Seidel, J.; Eliseev, E. A.; Morozovska, A. N.; Ramesh, R.; Pennycook, S. J.; et al. Mapping octahedral tilts and polarization across a domain wall in BiFeO3 from Z-contrast scanning transmission electron microscopy image atomic column shape analysis. *ACS Nano* **2010**, *4* (10), 6071-6079. DOI: 10.1021/nn1011539 From NLM PubMed-not-MEDLINE.

(18) O'Connell, E. N.; Moore, K.; McFall, E.; Hennessy, M.; Moynihan, E.; Bangert, U.; Conroy, M. TopoTEM: A python package for quantifying and visualizing scanning transmission electron microscopy data of polar topologies. *Microscopy and Microanalysis* **2022**, *28* (4), 1444-1452.

(19) Kim, Y. M.; Morozovska, A.; Eliseev, E.; Oxley, M. P.; Mishra, R.; Selbach, S. M.; Grande, T.; Pantelides, S. T.; Kalinin, S. V.; Borisevich, A. Y. Direct observation of ferroelectric field effect and vacancy-controlled screening at the BiFeO3/LaxSr1-xMnO3 interface. *Nat. Mater.* **2014**, *13* (11), 1019-1025, Article. DOI: 10.1038/nmat4058 From NLM PubMed-not-MEDLINE.

(20) Condurache, O.; Dražić, G.; Sakamoto, N.; Rojac, T.; Benčan, A. Atomically resolved structure of step-like uncharged and charged domain walls in polycrystalline BiFeO3. *Journal of Applied Physics* **2021**, *129* (5).

(21) Chen, S.; Yuan, S.; Hou, Z.; Tang, Y.; Zhang, J.; Wang, T.; Li, K.; Zhao, W.; Liu, X.; Chen, L. Recent progress on topological structures in ferroic thin films and heterostructures. *Advanced Materials* **2021**, *33* (6), 2000857.

(22) Hohenberger, S.; Jochum, J. K.; Van Bael, M. J.; Temst, K.; Patzig, C.; Höche, T.; Grundmann, M.; Lorenz, M. Enhanced magnetoelectric coupling in BaTiO3-BiFeO3 multilayers—an interface effect. *Materials* **2020**, *13* (1), 197.

(23) Fu, Z.; Chen, H.; Liu, Y.; Liu, M.; Liu, W.-M. Interface-induced ferroelectric domains and charged domain walls in Bi Fe O 3/Sr Ti O 3 superlattices. *Physical Review B* **2021**, *103* (19), 195301.

(24) Borisevich, A. Y.; Eliseev, E. A.; Morozovska, A. N.; Cheng, C. J.; Lin, J. Y.; Chu, Y. H.; Kan, D.; Takeuchi, I.; Nagarajan, V.; Kalinin, S. V. Atomic-scale evolution of modulated phases at the ferroelectric-antiferroelectric morphotropic phase boundary controlled by flexoelectric interaction. *Nat Commun* **2012**, *3*, 775, Article. DOI: 10.1038/ncomms1778 From NLM PubMed-not-MEDLINE.

(25) Gregg, J. Exotic domain states in ferroelectrics: searching for vortices and skyrmions. *Ferroelectrics* **2012**, *433* (1), 74-87.

(26) Das, S.; Hong, Z.; McCarter, M.; Shafer, P.; Shao, Y.-T.; Muller, D.; Martin, L.; Ramesh, R. A new era in ferroelectrics. *APL Materials* **2020**, *8* (12).

(27) Moore, K.; Bangert, U.; Conroy, M. Aberration corrected STEM techniques to investigate polarization in ferroelectric domain walls and vortices. *APL Materials* **2021**, *9* (2).

(28) Borisevich, A. Y.; Morozovska, A. N.; Kim, Y. M.; Leonard, D.; Oxley, M. P.; Biegalski, M. D.; Eliseev, E. A.; Kalinin, S. V. Exploring mesoscopic physics of vacancy-ordered systems through atomic scale observations of topological defects. *Phys. Rev. Lett.* **2012**, *109* (6), 065702, Article. DOI: 10.1103/PhysRevLett.109.065702 From NLM PubMed-not-MEDLINE.

(29) Li, Q.; Nelson, C. T.; Hsu, S. L.; Damodaran, A. R.; Li, L. L.; Yadav, A. K.; McCarter, M.; Martin, L. W.; Ramesh, R.; Kalinin, S. V. Quantification of flexoelectricity in PbTiO(3)/SrTiO(3) superlattice polar vortices using machine learning and phase-field modeling. *Nat Commun* **2017**, *8* (1), 1468, Article. DOI: 10.1038/s41467-017-01733-8 From NLM PubMed-not-MEDLINE.

(30) Nelson, C. T.; Vasudevan, R. K.; Zhang, X.; Ziatdinov, M.; Eliseev, E. A.; Takeuchi, I.; Morozovska, A. N.; Kalinin, S. V. Exploring physics of ferroelectric domain walls via Bayesian analysis of atomically resolved STEM data. *Nature communications* **2020**, *11* (1), 6361.

(31) Mukherjee, D.; Roccapriore, K. M.; Al-Najjar, A.; Ghosh, A.; Hinkle, J. D.; Lupini, A. R.; Vasudevan, R. K.; Kalinin, S. V.; Ovchinnikova, O. S.; Ziatdinov, M. A. A roadmap for edge computing enabled automated multidimensional transmission electron microscopy. *Microscopy Today* **2022**, *30* (6), 10-19.

(32) Belianinov, A.; He, Q.; Kravchenko, M.; Jesse, S.; Borisevich, A.; Kalinin, S. V. Identification of phases, symmetries and defects through local crystallography. *Nat Commun* **2015**, *6*, 7801, Article. DOI: 10.1038/ncomms8801 From NLM PubMed-not-MEDLINE.

(33) Kalinin, S. V.; Oxley, M. P.; Valleti, M.; Zhang, J. J.; Hermann, R. P.; Zheng, H.; Zhang, W. R.; Eres, G.; Vasudevan, R. K.; Ziatdinov, M. Deep Bayesian local crystallography. *Npj Computational Materials* **2021**, *7* (1), 12, Article. DOI: ARTN 181

10.1038/s41524-021-00621-6.

(34) Kervrann, C.; Boulanger, J. Optimal spatial adaptation for patch-based image denoising. *IEEE Transactions on Image Processing* **2006**, *15* (10), 2866-2878.

(35) Valleti, S.; Kalinin, S. V.; Nelson, C. T.; Peters, J. J.; Dong, W.; Beanland, R.; Zhang, X.; Takeuchi, I.; Ziatdinov, M. Unsupervised learning of ferroic variants from atomically resolved STEM images. *AIP Advances* **2022**, *12* (10).

(36) Jia, W.; Sun, M.; Lian, J.; Hou, S. Feature dimensionality reduction: a review. *Complex & Intelligent Systems* **2022**, *8* (3), 2663-2693.

(37) Meier, D.; Selbach, S. M. Ferroelectric domain walls for nanotechnology. *Nature Reviews Materials* **2022**, *7* (3), 157-173.

(38) Ishibashi, Y.; Takagi, Y. Note on ferroelectric domain switching. *Journal of the Physical Society of Japan* **1971**, *31* (2), 506-510.

(39) Li, L.; Xie, L.; Pan, X. Real-time studies of ferroelectric domain switching: a review. *Reports on Progress in Physics* **2019**, *82* (12), 126502.

(40) Davis, M.; Damjanovic, D.; Setter, N. Electric-field-, temperature-, and stress-induced phase transitions in relaxor ferroelectric single crystals. *Physical Review B—Condensed Matter and Materials Physics* **2006**, *73* (1), 014115.

(41) Lu, C.; Wu, M.; Lin, L.; Liu, J.-M. Single-phase multiferroics: new materials, phenomena, and physics. *National Science Review* **2019**, *6* (4), 653-668.

(42) Kusainova, A. M.; Lightfoot, P.; Zhou, W.; Stefanovich, S. Y.; Mosunov, A. V.; Dolgikh, V. A. Ferroelectric properties and crystal structure of the layered intergrowth phase Bi3Pb2Nb2O11Cl. *Chemistry of materials* **2001**, *13* (12), 4731-4737.

(43) Yi, Z.; Li, Y.; Liu, Y. Ferroelectric and piezoelectric properties of Aurivillius phase intergrowth ferroelectrics and the underlying materials design. *physica status solidi (a)* **2011**, *208* (5), 1035-1040.

(44) Nataf, G.; Guennou, M.; Gregg, J.; Meier, D.; Hlinka, J.; Salje, E.; Kreisel, J. Domain-wall engineering and topological defects in ferroelectric and ferroelastic materials. *Nature Reviews Physics* **2020**, *2* (11), 634-648.

(45) Seidel, J. Topological structures in ferroic materials. *Switzerland: Springer International Publishing* **2016**.

(46) Sifakis, E. G.; Prentza, A.; Koutsouris, D.; Chatziioannou, A. A. Evaluating the effect of various background correction methods regarding noise reduction, in two-channel microarray data. *Computers in biology and medicine* **2012**, *42* (1), 19-29.

(47) Chen, S. F.; Rosenfeld, R. A survey of smoothing techniques for ME models. *IEEE transactions on Speech and Audio Processing* **2000**, *8* (1), 37-50.

(48) Van Der Maaten, L.; Postma, E. O.; Van Den Herik, H. J. Dimensionality reduction: A comparative review. *Journal of machine learning research* **2009**, *10* (66-71), 13.

(49) Kang, S.; Park, J.; Lee, M. Machine learning-enabled autonomous operation for atomic force microscopes. *Review of Scientific Instruments* **2023**, *94* (12).

(50) Barakati, K.; Yuan, H.; Goyal, A.; Kalinin, S. V. Physics-based reward driven image analysis in microscopy. *arXiv preprint arXiv:2404.14146* **2024**.

(51) Barakati, K.; Pratiush, U.; Houston, A. C.; Duscher, G.; Kalinin, S. V. Unsupervised Reward-Driven Image Segmentation in Automated Scanning Transmission Electron Microscopy Experiments. *arXiv preprint arXiv:2409.12462* **2024**.

(52) Volkert, C. A.; Minor, A. M. Focused ion beam microscopy and micromachining. *MRS bulletin* **2007**, *32* (5), 389-399.

(53) Damjanovic, D. A morphotropic phase boundary system based on polarization rotation and polarization extension. *Applied Physics Letters* **2010**, *97* (6).

(54) Ziatdinov, M.; Nelson, C.; Vasudevan, R. K.; Chen, D.; Kalinin, S. V. Building ferroelectric from the bottom up: The machine learning analysis of the atomic-scale ferroelectric distortions. *Applied Physics Letters* **2019**, *115* (5).
(55) Reynolds, D. A. Gaussian mixture models. *Encyclopedia of biometrics* **2009**, *741* (659-663).
(56) Wan, H.; Wang, H.; Scotney, B.; Liu, J. A novel Gaussian mixture model for classification. In *2019 IEEE International Conference on Systems, Man and Cybernetics (SMC)*, 2019; IEEE: pp 3298-3303.
(57) Gopinath, R. A. Maximum likelihood modeling with Gaussian distributions for classification. In *Proceedings of the 1998 IEEE International Conference on Acoustics, Speech and Signal Processing, ICASSP'98 (Cat. No. 98CH36181)*, 1998; IEEE: Vol. 2, pp 661-664.
(58) Ishikawa, Y.; Kasai, S.; Aoki, Y.; Kataoka, H. Alleviating over-segmentation errors by detecting action boundaries. In *Proceedings of the IEEE/CVF winter conference on applications of computer vision*, 2021; pp 2322-2331.
(59) Likas, A.; Vlassis, N.; Verbeek, J. J. The global k-means clustering algorithm. *Pattern recognition* **2003**, *36* (2), 451-461.
(60) Sinaga, K. P.; Yang, M.-S. Unsupervised K-means clustering algorithm. *IEEE access* **2020**, *8*, 80716-80727.
(61) Asuero, A. G.; Sayago, A.; González, A. The correlation coefficient: An overview. *Critical reviews in analytical chemistry* **2006**, *36* (1), 41-59.
(62) Gao, P.; Kumamoto, A.; Ishikawa, R.; Lugg, N.; Shibata, N.; Ikuhara, Y. Picometer-scale atom position analysis in annular bright-field STEM imaging. *Ultramicroscopy* **2018**, *184*, 177-187.
(63) Gedraite, E. S.; Hadad, M. Investigation on the effect of a Gaussian Blur in image filtering and segmentation. In *Proceedings ELMAR-2011*, 2011; IEEE: pp 393-396.
(64) Ziou, D.; Tabbone, S. Edge detection techniques-an overview. *Распознавание образов и анализ изображен/Pattern Recognition and Image Analysis: Advances in Mathematical Theory and Applications* **1998**, *8* (4), 537-559.
(65) Illingworth, J.; Kittler, J. A survey of the Hough transform. *Computer vision, graphics, and image processing* **1988**, *44* (1), 87-116.
(66) Illingworth, J.; Kittler, J. The adaptive Hough transform. *IEEE Transactions on Pattern Analysis and Machine Intelligence* **1987**, (5), 690-698.
(67) Catalan, G.; Seidel, J.; Ramesh, R.; Scott, J. F. Domain wall nanoelectronics. *Reviews of Modern Physics* **2012**, *84* (1), 119-156.
(68) Gower, J. C. Properties of Euclidean and non-Euclidean distance matrices. *Linear algebra and its applications* **1985**, *67*, 81-97.
(69) Doersch, C. Tutorial on variational autoencoders. *arXiv preprint arXiv:1606.05908* **2016**.
(70) Kusner, M. J.; Paige, B.; Hernández-Lobato, J. M. Grammar variational autoencoder. In *International conference on machine learning*, 2017; PMLR: pp 1945-1954.
(71) Ziatdinov, M. A. *pyroVED*. https://github.com/ziatdinovmax/pyroVED (accessed.
(72) Valleti, M.; Ziatdinov, M.; Liu, Y.; Kalinin, S. V. Physics and chemistry from parsimonious representations: image analysis via invariant variational autoencoders. *npj Computational Materials* **2024**, *10* (1), 183.
(73) Węglarczyk, S. Kernel density estimation and its application. In *ITM web of conferences*, 2018; EDP Sciences: Vol. 23, p 00037.
(74) Sezan, M. I. A peak detection algorithm and its application to histogram-based image data reduction. *Computer vision, graphics, and image processing* **1990**, *49* (1), 36-51.